\documentstyle{article}

\def\etal{{et al.\,\,}}

\def\sec{^{\prime\prime}}
\def\min{^{\prime}}
\def\deg{^{\rm o}}

\def\msol{M_{\odot}}

\def\Hone{H{\sc i}}

\def\bootes{Bo\"{o}tes }

\begin{document}
\title{An \Hone\ survey of the Bo\"{o}tes void. I. The Data}
\author{Arpad Szomoru \thanks{Kapteyn Astronomical Institute, P.O.  Box
800, NL 9700 AV Groningen, The Netherlands, Visiting Astronomer, Kitt
Peak National Observatory, National Optical Astronomy Observatory} \and
J.  H.  van Gorkom \thanks{Columbia University, 538 W 120${\rm th}$
Street, New York NY 10027, USA} \and Michael D.  Gregg \thanks{Institute
of Geophysics and Planetary Physics, Lawrence Livermore National
Laboratory, P.O.  Box 808, L-413 Livermore, CA 94551-9900, USA, Visiting
Astronomer, Kitt Peak National Observatory, National Optical Astronomy
Observatory}}
\date{ }

\maketitle

\begin{abstract}
We present the results of a neutral hydrogen survey of
the \bootes void carried out with the VLA \footnote{The NRAO is operated
by Associated Universities, Inc., under a cooperative agreement with the
National Science Foundation.} (Napier \etal 1983) in D-array.  The
survey covers $\sim 1100$ Mpc$^{3}$, about 1\% of the volume of the void
as defined by Kirshner \etal 1987.  We observed 24 fields, centered on
known void galaxies; 16 of these were detected in \Hone.  Eighteen
uncataloged companion galaxies were discovered directly in the \Hone\
line at distances of $45\sec$ to $14.5\min$ from the target galaxies.
We also present the results of follow-up optical imaging observations
and discovery of one additional \bootes void galaxy, found through
spectroscopy of a number of apparent companions to known void members.
Our angular resolution is $\sim 1\min$ (45 kpc) \footnote{Throughout
this paper, we have assumed ${H_{0}}$ = ${\rm
100\,km\,s^{-1}Mpc^{-1}}$.}, each field has a size of $\sim 1\deg$ (2.7
Mpc).  The detected \Hone\ masses range from $8\times 10^{8}$ to
$1\times 10^{10} \msol$.  Typically our $2\sigma$ \Hone\ column density
sensitivity is $2\times 10^{19} {\rm cm^{-2}}$.  The radio and optical
data are analyzed and discussed in the following companion article
(Paper~2, Szomoru, van Gorkom, Gregg and Strauss 1996).

\end{abstract}

\end{document}